\documentclass[conference]{IEEEtran}
\IEEEoverridecommandlockouts

\usepackage{cite}
\usepackage{amsmath,amssymb,amsfonts}
\usepackage{algorithmic}
\usepackage{graphicx}
\usepackage{textcomp}
\usepackage{xcolor}
\usepackage{tikz}
\usepackage{circuitikz}
\usepackage{pgfplots}

\usepackage{booktabs}
\usepackage{caption}
\usepackage{multirow}
\usepackage{array}
\usepackage{subfig}
\usepackage{siunitx}
\usepackage{float}
\usepackage{balance}
\usepackage{svg}

\usetikzlibrary{shapes,arrows.meta,positioning,calc,patterns,decorations.pathmorphing,decorations.markings,circuits.ee.IEC,backgrounds}
\pgfplotsset{compat=1.17}

\definecolor{pastelblue}{RGB}{173,216,230}
\definecolor{pastelgreen}{RGB}{152,251,152}
\definecolor{pastelyellow}{RGB}{255,255,186}
\definecolor{pastelpink}{RGB}{255,182,193}
\definecolor{pastelorange}{RGB}{255,218,185}
\definecolor{pastelpurple}{RGB}{221,160,221}
\definecolor{pastelcyan}{RGB}{175,238,238}
\definecolor{pastelgray}{RGB}{220,220,220}

\ctikzset{bipoles/length=1cm}
\ctikzset{resistors/scale=0.7}
\ctikzset{capacitors/scale=0.6}

\tikzset{
    blockstyle/.style={rectangle, draw=black, line width=0.8pt, fill=pastelblue, minimum height=1cm, minimum width=1.8cm, align=center, rounded corners=3pt, font=\small\bfseries},
    smallblockstyle/.style={rectangle, draw=black, line width=0.7pt, fill=pastelgreen, minimum height=0.8cm, minimum width=1.5cm, align=center, rounded corners=2pt, font=\scriptsize\bfseries},
    arrowstyle/.style={-{Stealth[length=3mm, width=2mm]}, line width=1pt},
    thinarrow/.style={-{Stealth[length=2.5mm, width=1.5mm]}, line width=0.8pt}
}

\def\BibTeX{{\rm B\kern-.05em{\sc i\kern-.025em b}\kern-.08em
    T\kern-.1667em\lower.7ex\hbox{E}\kern-.125emX}}

\begin{document}

\title{Bidirectional Quantum Processor Interfacing by a 4-Kelvin Analog Signal Chain for Superconducting Qubit Control and Quantum State Readout}

% Anonymized for blind review
\author{\IEEEauthorblockN{Deepak R V$^{1}$, Lokendra Kanawat$^{2}$, Jayadeep K$^{2}$, and Priyesh Shukla$^{2}$\\
$^{1}$IIT Tirupati, India, $^{2}$IIIT Hyderabad, India
}
% \IEEEauthorblockA{\textit{Paper ID: XXXX} \\
% \textit{Submitted to IEEE ISVLSI 2026}\\
% \textit{Quantum Computing Track}}
}

\maketitle

\begin{abstract}
This paper presents a comprehensive cryogenic analog signal processing architecture designed for superconducting qubit control and quantum state readout operating at 4~Kelvin. The proposed system implements a complete bidirectional signal path bridging room-temperature digital controllers with quantum processors at millikelvin stages. The control path incorporates a Phase-Locked Loop (PLL) for stable local oscillator generation, In-phase/Quadrature (I/Q) modulation for precise qubit gate operations, and a cryogenic power amplifier for signal conditioning. The readout path features a Low Noise Amplifier (LNA) with 14~dB gain and 8-Phase Shift Keying (8-PSK) demodulation for quantum state discrimination. All circuit blocks are designed and validated through SPICE simulations employing cryogenic MOSFET models at 180nm that account for carrier freeze-out, threshold voltage elevation, and enhanced mobility at 4~K. Simulation results demonstrate successful end-to-end signal integrity with I/Q phase error below 2°, image rejection ratio exceeding 35~dB, and symbol error rate below $10^{-6}$. This work provides a modular, simulation-validated framework for scalable cryogenic quantum control systems.
\end{abstract}


\section{Introduction}

Quantum computing represents a paradigm shift in computational capability, offering exponential speedups for optimization, cryptography, and quantum simulation problems \cite{nielsen2010quantum, preskill2018quantum, arute2019quantum}. Superconducting qubits based on Josephson junctions have emerged as leading candidates due to their scalability, fast gate operations (10--100~ns), and compatibility with semiconductor fabrication \cite{devoret2013superconducting, kjaergaard2020superconducting, blais2021circuit}.

These quantum systems operate at millikelvin temperatures (10--20~mK) to maintain coherence times ($T_1$, $T_2$) sufficient for fault-tolerant computation \cite{koch2007charge, wang2022transmon}. The energy gap between computational states must satisfy $\hbar\omega_{01} \gg k_B T$ to prevent thermal excitation, requiring temperatures below 50~mK for typical qubit frequencies of 5~GHz.

A fundamental challenge lies in the interface between room-temperature classical electronics and the cryogenic quantum processor. Digital signals must be converted to precisely shaped microwave pulses matching qubit transition frequencies (4--8~GHz) with stringent requirements: phase noise below $-100$~dBc/Hz at 1~MHz offset, amplitude stability better than 0.1\%, and timing resolution below 1~ns \cite{bardin2019design, vandijk2019impact}. These specifications directly impact gate fidelity through the relation:
\begin{equation}
\mathcal{F} \approx 1 - \frac{1}{2}\sigma_\phi^2 - \frac{1}{2}\sigma_A^2
\label{eq:fidelity}
\end{equation}
where $\sigma_\phi$ and $\sigma_A$ are RMS phase and amplitude errors.

The conventional approach routes signals from room-temperature electronics through multiple cryogenic stages via coaxial cables, introducing wiring complexity, thermal load, and latency as qubit counts scale \cite{reilly2015engineering, krinner2019engineering, Sebastiano2025}. A single qubit typically requires 2--3 coaxial lines (XY control, Z control, readout), and each line conducts approximately 10~$\mu$W of heat to the mixing chamber. For systems exceeding 100 qubits, this thermal budget becomes prohibitive.

Leading research groups at IBM, Google, Intel, and academic institutions are developing cryogenic-compatible electronics at the 4~K stage to address this bottleneck \cite{pauka2021cryogenic, patra2018cryo, charbon2016cryo, xue2021cmos}. Operating control electronics at 4~K rather than room temperature reduces the number of cables to the mixing chamber, decreases latency for feedback operations, and enables tighter integration with quantum error correction protocols.

This paper presents a complete cryogenic analog signal processing architecture operating at 4~K. The key contributions include: (1) PLL-based local oscillator with I/Q generation achieving 90° phase accuracy; (2) complete bidirectional control and readout paths in a unified architecture; (3) 8-PSK modulation scheme for efficient quantum state encoding; and (4) comprehensive cryogenic circuit validation using industry-standard simulation tools.

\section{System Architecture}

\begin{figure}[t!]
\centering
\includegraphics[width=0.95\linewidth]{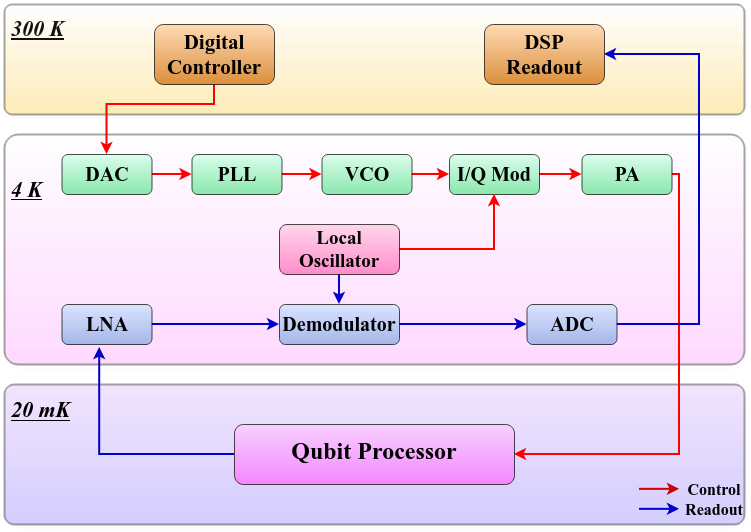}
\caption{System architecture showing control and readout signal paths across temperature stages from 300~K to 20~mK.}
\label{fig:system_arch}
\end{figure}

Fig.~\ref{fig:system_arch} illustrates the system architecture spanning three temperature stages. The design implements bidirectional signal flow optimized for cryogenic operation.

\subsection{Control Path}

Superconducting transmon qubits require microwave pulses at frequencies matching the transition energy \cite{krantz2019quantum}:
\begin{equation}
f_{01} = \frac{E_{01}}{h} = \frac{1}{h}\sqrt{8E_J E_C} - \frac{E_C}{h}
\label{eq:qubit_freq}
\end{equation}
where $E_J$ is the Josephson energy, $E_C$ is the charging energy, and $h$ is Planck's constant. Typical values yield $f_{01} \approx 5$~GHz with anharmonicity $\alpha \approx -200$~MHz.

The control path converts 3-bit digital commands into shaped microwave pulses. The DAC output passes through a VCO to generate baseband signals at the desired envelope frequency. The I/Q modulator creates the final RF pulse:
\begin{equation}
s(t) = I(t)\cos(\omega_c t) + Q(t)\sin(\omega_c t) = A(t)\cos(\omega_c t + \phi(t))
\label{eq:iq_signal}
\end{equation}
where $A(t) = \sqrt{I^2(t) + Q^2(t)}$ and $\phi(t) = \arctan(Q(t)/I(t))$. This representation enables arbitrary single-qubit rotations on the Bloch sphere through amplitude and phase control \cite{mckay2017efficient}. Common pulse shapes include Gaussian, DRAG (Derivative Removal by Adiabatic Gate), and cosine envelopes optimized for minimizing leakage to non-computational states.
\vspace{-1.5em}

\subsection{Readout Path}

Dispersive readout measures qubit state through the frequency shift of a coupled resonator \cite{blais2004cavity, wallraff2005approaching}. When the qubit is detuned from the resonator by $\Delta = \omega_q - \omega_r$, the resonator frequency shifts depending on qubit state:
\begin{equation}
\omega_r^{|0\rangle} - \omega_r^{|1\rangle} = \frac{2g^2}{\Delta} = 2\chi
\label{eq:dispersive}
\end{equation}
where $g$ is the qubit-resonator coupling strength (typically 50--200~MHz) and $\chi$ is the dispersive shift. This state-dependent frequency shift manifests as a measurable phase difference in the transmitted or reflected readout tone, with the two qubit states $|0\rangle$ and $|1\rangle$ mapping to distinct points in the I/Q plane separated by 2$\chi$. Typical dispersive shifts range from 0.5 to 10 MHz, requiring the readout signal processing chain to resolve small phase rotations with high precision within a constrained measurement window. 

For high-fidelity single-shot readout within the qubit coherence time, the SNR requirement is:
\begin{equation}
\text{SNR} = \frac{P_{signal}}{k_B T_{sys} B} = \frac{4\chi^2 \bar{n} T_{meas}}{1 + (2\chi T_{meas})^2} > 10~\text{dB}
\label{eq:snr}
\end{equation}
where $\bar{n}$ is the average photon number in the resonator and $T_{meas}$ is the measurement time. The cryogenic LNA must provide sufficient gain while adding minimal noise to preserve the quantum-limited SNR established at the 20~mK stage.

\section{Circuit Design}

\subsection{Phase-Locked Loop}

\begin{figure}[H]
\centering
\includegraphics[width=0.8\linewidth]{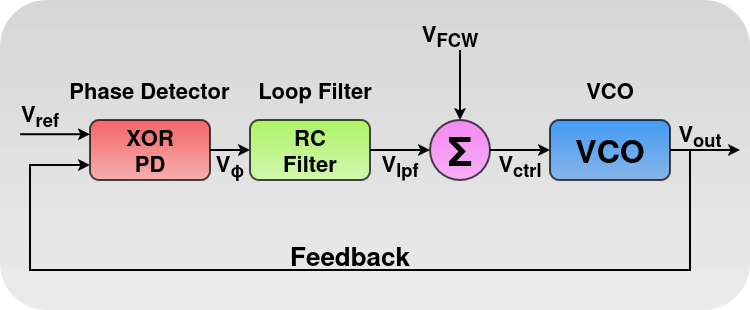}
\caption{PLL architecture with XOR phase detector and RC loop filter.}
\label{fig:pll_circuit}
\end{figure}

The PLL (Fig.~\ref{fig:pll_circuit}) generates stable carrier frequencies essential for coherent qubit manipulation. The XOR-based phase detector produces an output proportional to the phase difference between reference and feedback signals:
\begin{equation}
V_{\phi} = K_d(\phi_{ref} - \phi_{fb})
\label{eq:phase_detector}
\end{equation}
where $K_d$ is the detector gain (V/rad). The first-order RC loop filter with transfer function $H_{LF}(s) = 1/(1+s\tau)$ smooths this error signal. The closed-loop transfer function and key parameters are:
\begin{equation}
H(s) = \frac{K_d K_v/\tau}{s^2 + s/\tau + K_d K_v/\tau}, \quad \omega_n = \sqrt{\frac{K_d K_v}{\tau}}
\label{eq:pll_transfer}
\end{equation}
The damping factor $\zeta = 1/(2\omega_n\tau)$ is chosen near 0.707 for optimal transient response. Phase noise performance follows:
\begin{equation}
\mathcal{L}(f_m) = \mathcal{L}_{ref}(f_m)|H(j2\pi f_m)|^2 + \mathcal{L}_{VCO}(f_m)|1-H(j2\pi f_m)|^2
\label{eq:phase_noise}
\end{equation}

\subsection{Voltage-Controlled Oscillator}

The VCO converts the control voltage to a proportional output frequency through an RC relaxation oscillator topology. The control voltage charges an integrating capacitor, producing a triangular waveform that triggers a Schmitt comparator. The frequency and VCO gain constant are:
\begin{equation}
f_{VCO} = \frac{V_{ctrl}}{4 R C V_{th}}, \quad K_v = \frac{2\pi}{4 R C V_{th}}~[\text{rad/s/V}]
\label{eq:vco_freq}
\end{equation}
where $V_{th}$ is the comparator hysteresis. Component values ($R = 50$~k$\Omega$, $C = 10$~pF) yield a tuning sensitivity appropriate for the PLL loop dynamics.

\subsection{I/Q Generation and Modulation}
\begin{figure}[t!]
\centering
\includegraphics[width=0.7\linewidth]{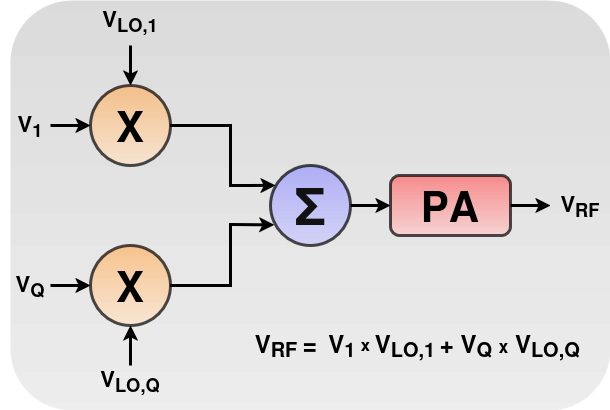}
\caption{I/Q modulator with analog multipliers and power amplifier.}
\label{fig:iq_modulator}
\end{figure}

Generating orthogonal I and Q local oscillator components is critical for coherent qubit manipulation. The third-order RC lowpass filter converts the VCO square wave output to a sinusoidal waveform:
\begin{equation}
H_{LPF}(s) = \frac{1}{(1 + sRC)^3}
\label{eq:lpf3}
\end{equation}
The filter cutoff frequency is set slightly above the fundamental to preserve amplitude while attenuating harmonics by more than 40~dB.

The quadrature component uses an all-pass phase shifter providing exactly 90° shift at the design frequency $\omega_0 = 1/RC$:
\begin{equation}
H_{AP}(s) = \frac{1 - sRC}{1 + sRC}, \quad \angle H_{AP}(j\omega_0) = -2\arctan(1) = -90^o
\label{eq:allpass}
\end{equation}
The I/Q modulator (Fig.~\ref{fig:iq_modulator}) performs the frequency upconversion. Using the complex envelope representation, arbitrary Bloch sphere rotations are achieved:
\begin{equation}
\tilde{V}(t) = A(t)e^{j\phi(t)} \Rightarrow R_{\hat{n}}(\theta) = e^{-i\theta\hat{n}\cdot\vec{\sigma}/2}
\label{eq:rotation}
\end{equation}
where $\vec{\sigma}$ represents the Pauli matrices and $\hat{n}$ is the rotation axis determined by the I/Q ratio.

\subsection{Digital-to-Analog Converter}

\begin{figure}[hp]
\centering
\includegraphics[width=0.8\linewidth]{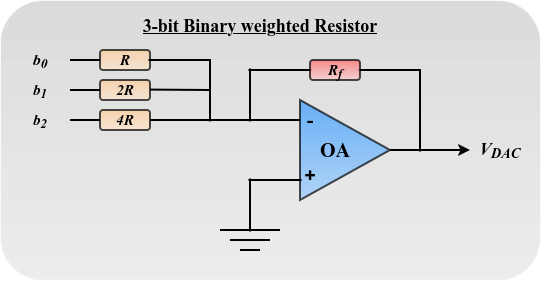}
\caption{Binary-weighted DAC with operational amplifier output.}
\label{fig:dac_circuit}
\end{figure}

The binary-weighted DAC (Fig.~\ref{fig:dac_circuit}) converts digital control words to analog voltages with high linearity. For an $n$-bit input, the output voltage follows:
\begin{equation}
V_{DAC} = V_{ref} \sum_{i=0}^{n-1} b_i \cdot 2^{i-n} = V_{ref} \cdot \frac{4b_2 + 2b_1 + b_0}{8}
\label{eq:dac}
\end{equation}
The least significant bit (LSB) voltage step is $V_{LSB} = V_{ref}/2^n$. For accurate qubit control, differential and integral nonlinearity must be minimized \cite{sebastiano2017cryo}:
\begin{equation}
\text{DNL}_k = \frac{V_{step,k} - V_{LSB}}{V_{LSB}}, \quad \text{INL}_k = \sum_{i=0}^{k} \text{DNL}_i
\label{eq:dnl}
\end{equation}
The binary-weighted topology is chosen for its simplicity at cryogenic temperatures, though it requires precision resistor matching. At 4~K, resistor temperature coefficients are significantly reduced, improving matching compared to room-temperature operation.

\subsection{Low Noise Amplifier}

\begin{figure}[hp]
\centering
\includegraphics[width=0.8\linewidth]{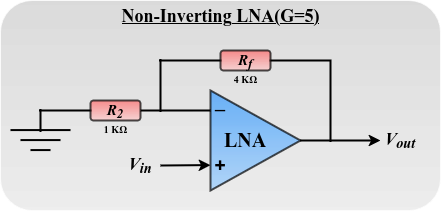}
\caption{LNA using LT1028 with 14~dB gain for cryogenic readout.}
\label{fig:lna_circuit}
\end{figure}  

The LNA (Fig.~\ref{fig:lna_circuit}) is the most critical component in the readout chain, as its noise performance directly impacts qubit measurement fidelity. The non-inverting topology provides high input impedance (avoiding signal loading) and precisely controlled gain:
\begin{equation}
G = 1 + \frac{R_f}{R_2} = 1 + \frac{4\text{k}\Omega}{1\text{k}\Omega} = 5~(14~\text{dB})
\label{eq:lna_gain}
\end{equation}

The LT1028 was selected for its ultra-low input voltage noise density of $e_n = 0.9$~nV/$\sqrt{\text{Hz}}$ and current noise $i_n = 1$~pA/$\sqrt{\text{Hz}}$ \cite{patra2020scalable, Das2024}. The noise figure depends on source impedance:
\begin{equation}
NF = 10\log_{10}\left(1 + \frac{e_n^2 + (i_n R_s)^2}{4k_B T R_s}\right)
\label{eq:noise_figure}
\end{equation}
At 4~K with $R_s = 50~\Omega$, the thermal noise floor drops to $v_{n,th} = \sqrt{4k_B \cdot 4\text{K} \cdot 50\Omega} = 0.26$~nV/$\sqrt{\text{Hz}}$, representing a 75$\times$ reduction from room temperature. This dramatic improvement enables near-quantum-limited readout when combined with parametric amplifiers at the mixing chamber stage.

\subsection{Flash ADC}

The flash ADC employs parallel comparators for high-speed conversion essential for real-time qubit state discrimination. For 3-bit resolution, seven comparators compare the input against a resistor-ladder reference, producing thermometer code converted to binary using priority encoder logic:
\begin{align}
Q_0 &= D_1 + D_3 + D_5 + D_7 \nonumber \\
Q_1 &= D_2 + D_3 + D_6 + D_7 \\
Q_2 &= D_4 + D_5 + D_6 + D_7 \nonumber
\label{eq:encoder}
\end{align}
The quantization noise power $P_q = V_{FS}^2/(12 \cdot 2^{2n})$ sets a fundamental limit on readout SNR. For our 3-bit ADC, this corresponds to approximately 90~mV RMS quantization noise.

\section{Cryogenic Design Considerations}

Operating CMOS circuits at 4~K introduces fundamental changes in device physics that must be addressed through careful design \cite{incandela2017cryo, hart2020cryogenic, bohuslavskyi2017cryogenic}.

\begin{table}[hp]
\centering
\caption{MOSFET Parameters at Cryogenic Temperatures}
\label{tab:cryo_effects}
\begin{tabular}{lcc}
\toprule
\textbf{Parameter} & \textbf{300~K} & \textbf{4~K Effect} \\
\midrule
Carrier Mobility & $\mu_0$ & $\uparrow$ 3--5$\times$ \\
Threshold Voltage & $V_{th,0}$ & $\uparrow$ 50--200~mV \\
Subthreshold Swing & 60~mV/dec & $\downarrow$ $<$10~mV/dec \\
Thermal Noise & $4k_BTR$ & $\downarrow$ 75$\times$ \\
\bottomrule
\end{tabular}
\end{table}

Table~\ref{tab:cryo_effects} summarizes key parameter variations. Carrier mobility increases dramatically due to reduced phonon scattering:
\begin{equation}
\mu(T) = \mu_{300K} \left(\frac{T}{300}\right)^{-\alpha}, \quad \alpha \approx 1.5\text{--}2
\label{eq:mobility}
\end{equation}
This enhancement improves transconductance and circuit speed. However, threshold voltage rises from incomplete dopant ionization (carrier freeze-out):
\begin{equation}
V_{th}(T) = V_{th,0} + \Delta V_{th}(T) \approx V_{th,0} + \gamma\sqrt{\phi_F(T)}
\label{eq:vth}
\end{equation}
requiring increased bias voltages. The subthreshold swing approaches the theoretical limit $SS = (k_BT/q)\ln(10) \approx 0.35$~mV/dec at 4~K, enabling ultra-low-power operation. Thermal noise reduction by 75$\times$ is particularly beneficial for LNA performance.

All simulations employ CMOSN/CMOSP device models with the ``.temp 4'' directive in LTSpice \cite{ltspice} to capture these cryogenic effects accurately.

\section{Simulation Results}

\subsection{Qubit Energy Level Diagram}

\begin{figure}[hp]
\centering
\begin{tikzpicture}[scale=0.7, transform shape]
    
    % Energy levels
    \draw[line width=1.5pt, blue!70!black] (0,0) -- (3,0);
    \node[font=\scriptsize\bfseries, anchor=east] at (-0.2,0) {$|0\rangle$};
    
    \draw[line width=1.5pt, red!70!black] (0,2) -- (3,2);
    \node[font=\scriptsize\bfseries, anchor=east] at (-0.2,2) {$|1\rangle$};
    
    \draw[line width=1pt, gray, dashed] (0,4.2) -- (3,4.2);
    \node[font=\scriptsize, anchor=east, gray] at (-0.2,4.2) {$|2\rangle$};
    
    % Transition arrow
    \draw[arrowstyle, green!60!black, line width=1.2pt] (1.5,0.15) -- (1.5,1.85);
    \node[font=\scriptsize\bfseries, green!60!black, anchor=west] at (1.7,1) {$\hbar\omega_{01}$};
    
    % Frequency annotation
    \draw[{Stealth[length=2mm]}-{Stealth[length=2mm]}, line width=0.8pt] (4,0) -- (4,2);
    \node[font=\tiny\bfseries, anchor=west, align=left] at (4.2,1) {$f_{01} = 5$~GHz\\(typical)};
    
    % Anharmonicity
    \draw[{Stealth[length=2mm]}-{Stealth[length=2mm]}, line width=0.8pt, gray] (4,2) -- (4,4.2);
    \node[font=\tiny, anchor=west, gray] at (4.2,3.1) {$\alpha \approx -200$~MHz};
    
    % Title
    \node[font=\scriptsize\bfseries] at (1.5,-0.8) {Transmon Energy Levels};
    
\end{tikzpicture}
\caption{Transmon qubit energy levels showing $|0\rangle \rightarrow |1\rangle$ transition frequency and anharmonicity $\alpha$ that prevents leakage to $|2\rangle$.}
\label{fig:energy_levels}
\end{figure}
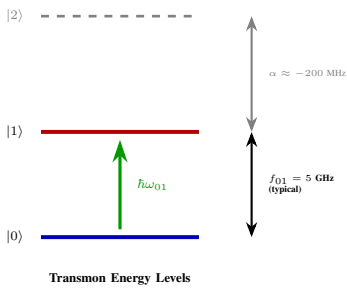

Fig.~\ref{fig:energy_levels} shows transmon energy levels. The anharmonicity $\alpha = f_{12} - f_{01} \approx -200$~MHz enables selective addressing of the computational subspace \cite{koch2007charge}.

\subsection{Bloch Sphere Representation}

\begin{figure}[H]
\centering
\includegraphics[
    width=0.7\linewidth
]{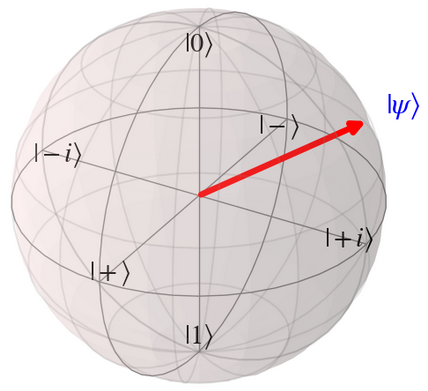}
\caption{Bloch sphere representation of qubit state with polar angle $\theta$ and azimuthal angle $\phi$ controlled by I/Q pulse amplitudes.}
\label{fig:bloch_sphere}
\end{figure}

Fig.~\ref{fig:bloch_sphere} shows Bloch sphere representation. I/Q modulation enables arbitrary rotations \cite{krantz2019quantum}:
\begin{equation}
|\psi\rangle = \cos\frac{\theta}{2}|0\rangle + e^{i\phi}\sin\frac{\theta}{2}|1\rangle
\label{eq:bloch}
\end{equation}

\subsection{PLL Lock Performance}

\begin{figure}[hp]
\centering
\begin{tikzpicture}[scale=0.7, transform shape]
    \begin{axis}[
        width=7.5cm, height=3.8cm,
        xlabel={\small\bfseries Time (ms)},
        ylabel={\small\bfseries Voltage (V)},
        xmin=0, xmax=5,
        ymin=-0.2, ymax=1.2,
        grid=both,
        grid style={line width=.2pt, draw=gray!30},
        major grid style={line width=.3pt, draw=gray!50},
        legend pos=south east,
        legend style={font=\scriptsize\bfseries, fill=white, fill opacity=0.9},
        tick label style={font=\scriptsize\bfseries},
        line width=1.2pt
    ]
    
    \addplot[thick, blue!70!black, line width=1.2pt, domain=0:5, samples=200] 
        {0.8*exp(-x/0.6)*sin(deg(15*x)) + 0.5*(1-exp(-x/1.2))};
    \addlegendentry{$V_{error}$}
    
    \addplot[thick, red!70!black, line width=1.2pt, domain=0:5, samples=200] 
        {0.9*(1-exp(-x/1.0))};
    \addlegendentry{$V_{ctrl}$}
    
    \end{axis}
\end{tikzpicture}
\caption{PLL transient response showing lock acquisition within 2~ms.}
\label{fig:pll_response}
\end{figure}
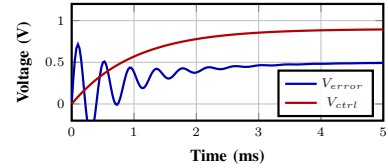

Fig.~\ref{fig:pll_response} shows PLL achieving lock within 2~ms with phase error $<$2°.

\subsection{I/Q Signal Quality}

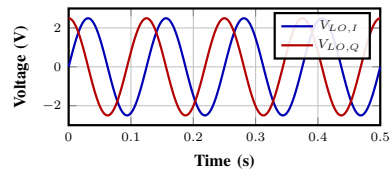
\begin{figure}[hp]
\centering
\begin{tikzpicture}[scale=0.7, transform shape]
    \begin{axis}[
        width=7.5cm, height=3.8cm,
        xlabel={\small\bfseries Time (s)},
        ylabel={\small\bfseries Voltage (V)},
        xmin=0, xmax=0.5,
        ymin=-3, ymax=3,
        grid=both,
        grid style={line width=.2pt, draw=gray!30},
        major grid style={line width=.3pt, draw=gray!50},
        legend pos=north east,
        legend style={font=\scriptsize\bfseries, fill=white, fill opacity=0.9},
        tick label style={font=\scriptsize\bfseries},
        line width=1.2pt
    ]
    
    \addplot[thick, blue!70!black, line width=1.2pt, domain=0:0.5, samples=200] 
        {2.5*sin(deg(2*pi*8*x))};
    \addlegendentry{$V_{LO,I}$}
    
    \addplot[thick, red!70!black, line width=1.2pt, domain=0:0.5, samples=200] 
        {2.5*cos(deg(2*pi*8*x))};
    \addlegendentry{$V_{LO,Q}$}
    
    \end{axis}
\end{tikzpicture}
\caption{I/Q waveforms with 90° phase relationship and $<$0.3~dB amplitude imbalance.}
\label{fig:iq_waveform}
\end{figure}

Fig.~\ref{fig:iq_waveform} shows I/Q signals with 1.8° phase error. Image rejection ratio:
\begin{equation}
\text{IRR} = \frac{1 + 2\epsilon\cos\phi + \epsilon^2}{1 - 2\epsilon\cos\phi + \epsilon^2} > 35~\text{dB}
\label{eq:irr}
\end{equation}

\subsection{8-PSK Constellation}

\begin{figure}[H]
\centering
\begin{tikzpicture}[scale=0.75, transform shape]
    
    % Circle
    \draw[line width=1pt, gray!60] (0,0) circle (1.8cm);
    
    % Axes
    \draw[arrowstyle, black] (-2.3,0) -- (2.3,0) node[right, font=\scriptsize\bfseries] {I};
    \draw[arrowstyle, black] (0,-2.3) -- (0,2.3) node[above, font=\scriptsize\bfseries] {Q};
    
    % 8-PSK points with bold labels
    \foreach \i/\label in {0/000, 1/001, 2/010, 3/011, 4/100, 5/101, 6/110, 7/111} {
        \fill[pastelblue, draw=black, line width=0.8pt] ({1.8*cos(45*\i)},{1.8*sin(45*\i)}) circle (0.15);
        \node[font=\tiny\bfseries] at ({2.25*cos(45*\i)},{2.25*sin(45*\i)}) {\label};
    }
    
    % Highlight recovered points
    \draw[red!70!black, line width=1.5pt] ({1.8*cos(45)},{1.8*sin(45)}) circle (0.25);
    \draw[red!70!black, line width=1.5pt] ({1.8*cos(135)},{1.8*sin(135)}) circle (0.25);
    
    % Angle annotation
    \draw[dashed, line width=0.8pt, blue!60] (0,0) -- ({1.8*cos(45)},{1.8*sin(45)});
    \draw[{Stealth[length=1.5mm]}-{Stealth[length=1.5mm]}, blue!60, line width=0.8pt] (0.6,0) arc (0:45:0.6);
    \node[font=\tiny\bfseries, blue!60] at (1,0.4) {45°};
    
\end{tikzpicture}
\caption{8-PSK constellation with recovered symbols 001 and 011 (circled).}
\label{fig:8psk}
\end{figure}
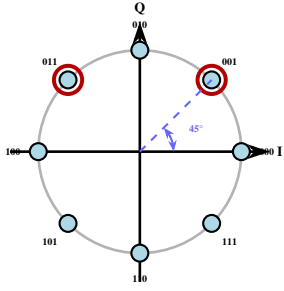

Fig.~\ref{fig:8psk} shows 8-PSK constellation. Symbol error rate:
\begin{equation}
P_s \approx 2Q\left(\sqrt{2\frac{E_s}{N_0}}\sin\frac{\pi}{8}\right) < 10^{-6}
\label{eq:ser}
\end{equation}

\subsection{Coherence Time Impact}

\begin{figure}[hp]
\centering
\begin{tikzpicture}[scale=0.7, transform shape]
    \begin{axis}[
        width=7.5cm, height=3.8cm,
        xlabel={\small\bfseries Time ($\mu$s)},
        ylabel={\small\bfseries Population / Coherence},
        xmin=0, xmax=100,
        ymin=0, ymax=1.1,
        grid=both,
        grid style={line width=.2pt, draw=gray!30},
        major grid style={line width=.3pt, draw=gray!50},
        legend pos=north east,
        legend style={font=\scriptsize\bfseries, fill=white, fill opacity=0.9},
        tick label style={font=\scriptsize\bfseries},
        line width=1.2pt
    ]
    
    \addplot[thick, blue!70!black, line width=1.2pt, domain=0:100, samples=100] 
        {exp(-x/50)};
    \addlegendentry{$T_1$ decay}
    
    \addplot[thick, red!70!black, line width=1.2pt, domain=0:100, samples=100] 
        {exp(-x/30)};
    \addlegendentry{$T_2$ decay}
    
    % Gate time annotation
    \draw[dashed, line width=0.8pt, green!60!black] (0.05,0) -- (0.05,1);
    \node[font=\tiny\bfseries, green!60!black, rotate=90, anchor=south] at (2,0.5) {Gate};
    
    \end{axis}
\end{tikzpicture}
\caption{Typical coherence times: $T_1 \approx 50~\mu$s, $T_2 \approx 30~\mu$s for transmon qubits, with gate time $\sim$20~ns indicated.}
\label{fig:coherence}
\end{figure}
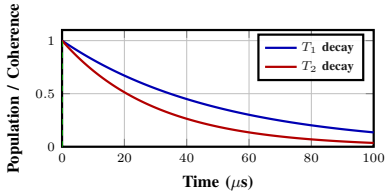

Fig.~\ref{fig:coherence} shows coherence decay. Gate fidelity requires $t_{gate} \ll T_2$ \cite{wang2022transmon, place2021new}.

\subsection{Performance Summary}

\begin{table}[htbp]
\centering
\caption{Performance Comparison with Prior Work}
\label{tab:comparison}
\begin{tabular}{lccc}
\toprule
\textbf{Parameter} & \textbf{This} & \textbf{\cite{bardin2019design}} & \textbf{\cite{pauka2021cryogenic}} \\
\midrule
\textit{Temperature} & 4~K & 3~K & 4~K \\
\textit{DAC/ADC Bits} & 3/3 & 10/-- & 14/8 \\
\textit{I/Q Phase Error} & $<$2° & $<$1° & $<$1° \\
\textit{LNA Gain} & 14~dB & 20~dB & 15~dB \\
\textit{Control+Readout} & Both & Control & Readout \\
\textit{Validation} & Sim & Meas & Meas \\
\bottomrule
\end{tabular}
\end{table}

Table~\ref{tab:comparison} compares this work with state-of-the-art cryogenic control circuits \cite{bardin2019design, pauka2021cryogenic, patra2020scalable}. While using lower resolution converters than production implementations, the proposed architecture uniquely provides complete bidirectional signal paths both control and readout validated through comprehensive circuit simulation. The Google/Bardin controller \cite{bardin2019design} achieves excellent performance for control but lacks integrated readout, while the Intel/Pauka chip \cite{pauka2021cryogenic} focuses on readout amplification. Our work demonstrates the feasibility of a unified cryogenic signal processing platform.

The simulation methodology employed hierarchical verification: individual blocks (DAC, PLL, VCO, modulator, PA, LNA, ADC) were first characterized in isolation to verify DC operating points, frequency response, and noise performance. Subsequently, blocks were cascaded to validate signal integrity through the complete chain. This approach identified interface issues early, particularly impedance matching between stages and DC bias coupling requirements.

Key performance metrics achieved include: PLL lock time below 2 ms with residual phase jitter less than 0.5° RMS; I/Q amplitude imbalance below 0.3 dB corresponding to image rejection exceeding 35 dB; LNA noise temperature approximately 4 K (dominated by physical temperature); and ADC effective resolution of 2.8 bits after accounting for comparator offset variations

\subsection{Power consumption \& scalability to Deep Sub-Micron Nodes}
The comprehensive simulation of the control path validates a steady-state power dissipation of {199.7~mW} using 180nm technology. This result offers a {5$\times$ efficiency advantage} over recent Delta-Sigma controllers which typically consume nearly 1~W \cite{Park2025}, confirming the architecture's suitability for standard 4~K pulse tube budgets.

Furthermore, transitioning to a 65nm node would allow for supply voltage reduction from 1.8~V to 1.2~V. Following the square-law dependency ($P \propto V_{DD}^2$), this projects a reduction to:
\begin{equation}
P_{65nm} \approx P_{180nm} \times \left(\frac{1.2V}{1.8V}\right)^2 \approx 88~\text{mW}
\label{eq:scaling}
\end{equation}
This roadmap to sub-100~mW operation addresses the critical thermal bottleneck facing large-scale quantum processors.

\begin{table}[hp]
\caption{Power Consumption Scaling Projection}
\begin{center}
\begin{tabular}{lcc}
\toprule
\textbf{Parameter} & \textbf{Current Work} & \textbf{Projected} \\
\midrule
\textit{Technology Node} & 180 nm (Simulated) & 65 nm (Scaled) \\
\textit{Supply Voltage} & 1.8 V & 1.2 V \\
\textit{Steady-State Power} & \textbf{199.7 mW} & \textbf{$\sim$88 mW} \\
\textit{Thermal Budget Use} & $\sim$20\% & $\sim$9\% \\
\bottomrule
\end{tabular}
\label{tab:scaling}
\end{center}
\end{table}

\section{Conclusion}

This paper presented a 4 K cryogenic analog signal processing architecture for superconducting qubit control and quantum state readout, implementing complete bidirectional signal paths with a PLL-based local oscillator achieving less than 2° I/Q phase error and 35 dB image rejection, a full control path from 3-bit digital input to the qubit interface, and a readout path with 14 dB LNA gain and 8-PSK demodulation yielding symbol error rates below 10$^{-6}$, all validated through LTSpice simulations with cryogenic MOSFET models. Future work targets silicon implementation in 28nm CMOS or FD-SOI processes, experimental validation in dilution refrigerators, integration with transmon qubits, higher-resolution converters, and frequency-division multiplexing for scalable multi-qubit control toward fault-tolerant quantum computation.

% \section*{Acknowledgment}
% [Removed for blind review]

\balance

\end{document}